\documentclass[10pt]{iopart}

\usepackage{iopams}
\usepackage{psfig}
\usepackage{citesort}
\usepackage{verbatim}

\begin{document}

%%%%%%%%%%%%%%%%%%%%%%%%%%%%%%%%%%%%%%%%%%%%%%%%%%%%%%%%%%%%%%%%%%%%%%%%%%%%%%%%%%%%%%%%%
% Journal identifier can be put here if required, e.g.
\jl{1}
%%%%%%%%%%%%%%%%%%%%%%%%%%%%%%%%%%%%%%%%%%%%%%%%%%%%%%%%%%%%%%%%%%%%%%%%%%%%%%%%%%%%%%%%%

\bibliographystyle{jpc}

%%%%%%%%%%%%%%%%%%%%%%%%%%%%%%%%%%%%%%%%%%%%%%%%%%%%%%%%%%%%%%%%%%%%%%%%%%%%%%%%%%%%%%%%%
%
%\title[Dispersion of temporally stable coherent states]{Dispersion of %
%Klauder's temporally stable coherent states for the hydrogen atom}
%
%%%%%%%%%%%%%%%%%%%%%%%%%%%%%%%%%%%%%%%%%%%%%%%%%%%%%%%%%%%%%%%%%%%%%%%%%%%%%%%%%%%%%%%%%

\letter{Dispersion of Klauder's temporally stable coherent states for the hydrogen atom}

\author{Paolo Bellomo\dag\footnote[3]{To
whom correspondence should be addressed.} and C. R. Stroud, Jr.\dag}

\address{\dag\ Rochester Theory Center for Optical Science and Engineering and
The Institute of Optics, University of Rochester, Rochester New York, 14627-0186, USA.}

\begin{abstract}

We study the dispersion of the ``temporally stable" coherent
states for the hydrogen atom introduced by Klauder. 
These are states which under temporal evolution
by the hydrogen atom Hamiltonian retain their coherence properties.
We show that in the hydrogen atom such wave packets do not move
quasi-classically; {\em i.e.}, they do not follow 
with no or little dispersion the 
Keplerian orbits of the classical electron. 
The poor quantum-classical correspondence does not improve
in the semiclassical limit.

\end{abstract}

\pacs{%
03.65.Ca, 32.80.Rm, 33.80.Rv
-- To appear in J. Phys. A
}

%\submitted
%\maketitle
%
%%%%%%%%%%%%%%%%%%%%%%%%%%%%%%%%%%%%%%%%%%%%%%%%%%%%%%%%%%%%%%%%%%%%%%%%%%%%%%%%%%%%%%%

\section{Temporally stable coherent states for the hydrogen atom}

In recent years accurate experimental results have generated 
renewed interest 
{\cite{carlos_86a,carlos_91a,nauenberg_89a,nauenberg_90a,%
nauenberg_94a,carlos_94a,eberly_94a,eberly_96b,%
delande_95a,delande_95b,turgay_96a,turgay_97a,bialynicki_97a}} 
in a long standing problem of quantum physics, which was 
addressed, albeit unsuccessfully, already in 1926 by Schr{\"o}dinger 
himself {\cite{schroedinger_26a}}: that is the problem of constructing localized, 
non-spreading wave packets in the hydrogen atom which travel along the 
classical trajectories (Keplerian ellipses) of the electron.
A partial solution to the problem has been found only in recent 
years {\cite{nauenberg_89a,nauenberg_90a,nauenberg_94a}}, 
in the form of wave packets which are superpositions of coherent states of the 
angular momentum {\cite{delande_89a}} and minimize quantum fluctuations,
confining the bound electron (for high quantum numbers) to a Keplerian ellipse.
Such wave packets, however, do not remain localized in the angular variable and
spread along the classical ellipse. Eventually they display interference 
fringes as the front of the packet catches up with its tail, and finally they also 
show quantum revivals and super-revivals 
{\cite{carlos_86a,nauenberg_89a,carlos_91a,carlos_97a,perelman_89}}.

Most recently, two alternative solutions to this fundamental problem
have been proposed by  Klauder {\cite{klauder_96a} and
also Majumdar and Sharatchandra {\cite{majumdar_97a}, who model the 
properties of their coherent states for the hydrogen atom
on the ones of the more familiar coherent states for the
harmonic oscillator {\cite{tannoudji_77a}}. 
More precisely, those authors construct states which enjoy the property of 
{\em temporal stability} {\cite{klauder_96a}. 
Under the action of the harmonic propagator
a coherent state for the harmonic oscillator remains a coherent state; 
by a judicious choice of the coefficients in the expansion over the 
eigenstates of the hydrogen atom Hamiltonian, Klauder {\cite{klauder_96a}} 
and also Majumdar and Sharatchandra {\cite{majumdar_97a}}
construct wave packets which satisfy exactly the same property 
(i.e., temporal stability) retaining their ``coherence" under the 
action of the hydrogenic propagator. 
This is done extending the 
parameter space of the wave packet from the 
polar coordinates  $r, \theta$ where 
$0 \leq r < \infty$, $- \pi < \theta \leq \pi$,
to their covering space, namely the domain 
$0 \leq r < \infty$, $- \infty < \theta < \infty$, 
with an appropriately modified measure, which is
identical to the classical phase space measure {\cite{klauder_96a,majumdar_97a}}.
In this new space the temporally stable coherent 
states are parameterized continuously and also
admit a resolution of unity (limited to the bound part of the spectrum), very much
like the coherent states for the harmonic oscillator.

However, in the case of the hydrogen atom 
the question of the relation between quasi-classical dynamics and 
temporal stability of coherent states remained unanswered.
The harmonic spectrum, with its equally spaced levels, occupies a very 
special place in physics (for example,  its spacings distribution 
is not Poissonian {\cite{eckhardt_88a}}, in contrast with 
most other integrable systems) 
and therefore it is not obvious that 
temporal stability is equivalent to 
quasi-classical dynamics also in
the case of the hydrogenic spectrum. 
In fact, in this paper we study precisely the dynamical properties of temporally stable
coherent states for the hydrogen atom, and we find
that they {\em do not move along 
classical Keplerian ellipses with no or little dispersion}.
We also demonstrate that in the semiclassical limit 
the poor correspondence between the dynamics of
temporally stable states and classical orbits does not improve. Therefore,
the well known wave packets 
which confine the electron on a Kepler 
ellipse {\cite{nauenberg_89a,nauenberg_90a,nauenberg_94a}}, but
not in the angular variable, remain the most accurate
solution to the problem of 
non-spreading wave packets in the hydrogen atom.

\section{The autocorrelation function}

The most natural way to study the dynamics of any quantum state is to 
solve the time dependent Schr{\"o}dinger equation and 
actually observe the time evolution of the 
state. In principle this is particularly 
simple if the projections of the state on the eigenstates of 
the Hamiltonian are known exactly. Unfortunately, in the hydrogen atom
each eigenenergy corresponds to a very large manifold of degenerate eigenstates, and the actual
calculation of the temporal evolution of a wave packet which spans several 
(in principle infinitely many) high energy eigenstates
becomes computationally overwhelming. 

However, there is an alternative way to extract information
about the dynamics of a wave packet $| \phi \rangle$, 
and that is to calculate its autocorrelation function,
which in atomic units (which we use throughout this paper) is defined as 
{\cite{heller_89a,nauenberg_90a}}:
\begin{equation}
\label{klauder_1}
	C(t) = \left| \langle \phi | \exp ^{ -i {\hat H} t} | \phi \rangle \right| ^{2}	\; .
\end{equation}
The autocorrelation function 
carries information about when and how a state returns to 
its original configuration; 
this is precisely the information necessary to determine whether temporal stability 
warrants  quasi-classical dynamics in the hydrogen atom as it does in the harmonic
oscillator. The classical dynamics of the hydrogen atom is obviously periodic and 
a normalized wave packet which travels along a Keplerian ellipse with 
no or little dispersion yields an
autocorrelation function which returns to one (i.e., its maximum value),
or very close to one, 
every Kepler period {\cite{nauenberg_90a}.
We study the autocorrelation function of temporally stable coherent states
(we discuss here the states proposed by Klauder {\cite{klauder_96a}}, and study 
the particulars of a similar proposal by 
Majumdar and Sharatchandra {\cite{majumdar_97a}} elsewhere {\cite{myself_10}}), and
we demonstrate that in the hydrogen atom temporal stability 
is not equivalent to quasi-classical dynamics.

The expansion of temporally stable 
states on the hydrogenic eigenbasis reads {\cite{klauder_96a}}:
\begin{equation}
\label{klauder_3}
	| s \ \gamma \ \bar{\Omega} \rangle = 
	{\rm e} ^{ - s^{2} / 2 } \sum_{ n=1 }^{ \infty } 
	\frac{ s^{n-1} {\rm e} ^{ i \gamma / n^{2} }  }{  \sqrt{ ( n-1 ) ! }  } 
        | n \ {\bar{\Omega}} \rangle \; ,
\end{equation}
where $s, \ \gamma$ and 
${\bar{\Omega}}=({\bar \theta}, {\bar \phi}, {\bar \psi})$ 
are continuous parameters and the states 
$| n \ {\bar{\Omega}} \rangle$ are defined as {\cite{klauder_96a}}:
\begin{equation}
	{\hspace{-2.3cm}}
	| n \ {\bar{\Omega}} \rangle =
	\! \sum_{ \ell = 0 }^{n-1} \sum_{ m = -\ell }^{ \ell }
	\left[ \frac{ ( 2 \ell + 1 ) ! }{ ( \ell + m ) ! ( \ell - m ) ! } \right]^{\frac{1}{2}}
	\!\! \left( \sin \frac{ \bar{\theta} }{2} \right) ^{\ell - m} 
	\!\! \left( \cos \frac{ \bar{\theta} }{2} \right) ^{\ell + m}
	\!\!\! {\rm e}^{i( m \bar{\phi} + 
	\ell \bar{\psi} ) } | n \ \ell \ m \rangle \, .
\label{klauder_4}
\end{equation}
The states $| n \ {\bar{\Omega}} \rangle$ are normalized 
to the degeneracy of the hydrogenic {\em n}-manifold:
\begin{equation}
	{\hspace{-2.0cm}}
	\langle n \ {\bar{\Omega}} | n^{\prime} \ {\bar{\Omega}} \rangle =
	\delta_{n,n^{\prime}}  
	\sum_{ \ell = 0 }^{n-1} ( 2 \ell + 1 ) \!\!
	\sum_{ m = -\ell }^{ \ell } 
	\frac{ ( 2 \ell ) ! }{ ( \ell + m ) ! ( \ell - m ) ! } 
	\left( \sin^{2} \frac{ \bar{\theta} }{2} \right) ^{\ell - m} \!\!\!\!
	\left( \cos^{2} \frac{ \bar{\theta} }{2} \right) ^{\ell + m} \!\!\!\! ,
\label{klauder_5}
\end{equation}
and setting $k = \ell + m$ one obtains:
\begin{eqnarray}
	{\hspace{-1.5cm}}
	\langle n \ {\bar{\Omega}} | n^{\prime} \ {\bar{\Omega}} \rangle
	& = &
	\delta_{n,n^{\prime}}  
	\sum_{ \ell = 0 }^{n-1} ( 2 \ell + 1 ) \!
	\sum_{ k = 0 }^{ 2 \ell } 
	\frac{ ( 2 \ell ) ! }{ k ! ( 2 \ell - k ) ! } 
	\left( \sin^{2} \frac{ \bar{\theta} }{2} \right) ^{ 2 \ell - k} \!\!
	\left( \cos^{2} \frac{ \bar{\theta} }{2} \right) ^{k} = \nonumber \\
	& = &
	\delta_{n,n^{\prime}}  
	\sum_{ \ell = 0 }^{n-1} ( 2 \ell + 1 ) \!
	\left( \sin^{2} \frac{ \bar{\theta} }{2} +
	\cos^{2} \frac{ \bar{\theta} }{2} \right) ^{2 \ell } \!\!\! = 
	n^{2} \, \delta_{n,n^{\prime}}  \; .
\label{klauder_6}
\end{eqnarray}
Therefore the normalization coefficient of the state in Eq. (\ref{klauder_3}) is:
\begin{equation}
	{\hspace{-2.cm}}
	\langle s \ \gamma \ \bar{\Omega} | s \ \gamma \ \bar{\Omega} \rangle 
	=
	{\rm e}^{-s^{2}}
	\sum_{n=1}^{\infty} 
	\frac{ n^{2} s^{ 2(n-1) } }{ (n-1)! } 
	=
	{\rm e} ^{ - s^{2} }	
	\left[ \frac{ \partial }{ \partial ( s^2 ) } s^{2} \right]^{2}
	{\rm e} ^{ s^{2} } \! = 
	( s^{2} + 1 ) ^{2} + s^{2} \; ,
\label{klauder_7}
\end{equation}
where the square of the operator between brackets indicates that the operator must be
applied twice.

We are now in the position to calculate the autocorrelation function of temporally
stable coherent states:
\begin{equation}
	{\hspace{-1.cm}}
	C(t)
	= 
	\left| \langle s \ \gamma \ \bar{\Omega} | 
	\exp ^{ -i {\hat H} t} 
	| s \ \gamma \ \bar{\Omega} \rangle \right| ^{2} = 
	\frac{ {\rm e}^{ - 2 s^{2} }
	\left|
	\sum_{n=1}^{\infty} 
	\frac{ n^{2} s^{ 2(n-1) } }{ (n-1)! } {\rm e} ^{ i t / 2 n^{2} }
	\right| ^{2}
	 }{ \left[ ( s^{2} + 1 )^{2} + s^{2} \right]^{2} } \; .
\label{klauder_8}
\end{equation}
The sum in Eq. (\ref{klauder_8}) can be evaluated numerically with great efficiency. 
Clearly, we evaluated a finite sum, however only relatively few terms in the sum of 
Eq. (\ref{klauder_8}) yield a nonnegligible contribution. We 
neglected only terms which are $< 10^{-100}$, so that
our results are an extremely accurate approximation.
We also tested our results by
repeating the calculations with increasingly stringent constraints on 
the terms to be neglected, and we did not notice any significant changes in the final results.
In Fig. 1 we plot our results for two states; 
in Fig. 1-(a) we show 
the autocorrelation function for a temporally stable state with
$s=5$, whereas in Fig. 1-(b) we set $s=20$. 
In both cases the dispersion of the wave
packet is very different from the one of a 
quasi-classical packet, moving periodically and with
no or little dispersion along a Kepler ellipse {\cite{nauenberg_90a}}. 
%
%%%\begin{comment}
%
%%%%%%%%%%%%%%%%%%%%%%%%%%%%%%%%%%%%%%%%%%%%%%%%%%%%%%%%%%%%%%%%%%%%%%%%%%%%%%%
%
% Importing a Figure
%
\begin{figure}
\centerline{\psfig{file=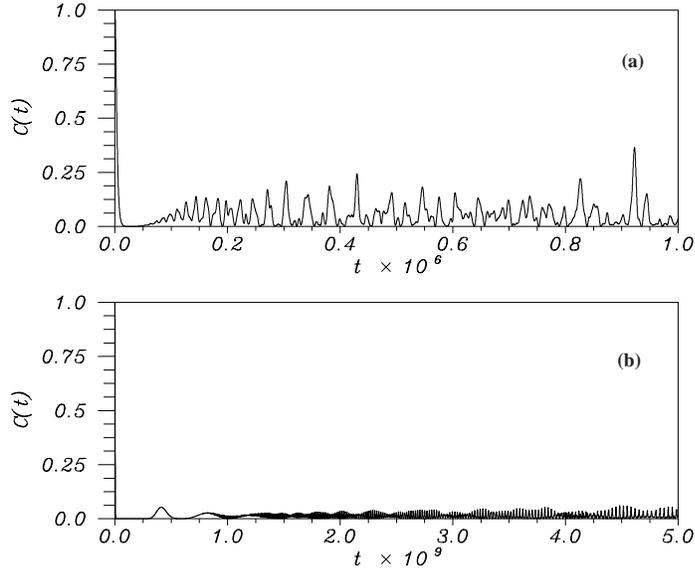,height=7.5cm,angle=0}}
\caption{
The autocorrelation function for temporally stable coherent states:
in Fig. (a) we set $s=5$ and in Fig. (b) we set $s=20$. 
The small recurrence peaks of Fig. (a) essentially disappear
in Fig. (b), i.e., for a wave packet with a much larger average
principal quantum number. \\
}
\label{k_fig_1}
\end{figure}
%
%%%%%%%%%%%%%%%%%%%%%%%%%%%%%%%%%%%%%%%%%%%%%%%%%%%%%%%%%%%%%%%%%%%%%%%%%%%%%%%
%
%%%\end{comment}
%
Surprisingly, the peaks of 
Fig. 1-(a), 
which imply at least some partial recurrences on part of the temporally
stable wave packet, disappear almost completely in Fig. 1-(b), where one
would expect the wave packet to
behave more classically, as we explain next.

The coefficients of the sum in Eq. (\ref{klauder_8}) are (including normalization):
\begin{equation}
	c_{n} = \frac{ n^{2} s^{ 2(n-1) } }{ (n-1)! \left[ ( s^{2} + 1 )^{2} + s^{2} \right] }
\label{klauder_8a}
\end{equation}
and are determined by the constraint that the 
integration over the parameter space
yields a resolution of the unity for the bound part of the spectrum {\cite{klauder_96a}}.
Although the $c_{n}$'s are not exactly Poissonian 
they still enjoy to a good approximation many properties of the Poisson distribution.
For example, they
satisfy the following recursion relation:
\begin{equation}
	c_{n+1} = \frac{ ( n+1 ) ^{2} }{ n^{3} } s^{2} c_{n} \; ,
\label{klauder_9}
\end{equation}
and the distribution is centered around a principal quantum number ${\bar n}$
such that:
\begin{equation}
	\frac{( {\bar n}+1 ) ^{2} }{ {\bar n}^{3} } s^{2} = 1 
	\Rightarrow {\bar n} \approx s^{2} \; .
\label{klauder_10}
\end{equation}
Therefore the results of Fig.'s 1-(a) and 1-(b) 
refer to wave packets in which the eigenstate
carrying the largest weight has 
principal quantum number ${\bar n} \approx 25$ 
and ${\bar n} \approx 400$ respectively.
Therefore one would expect a closer quantum-classical correspondence in 
the second wave packet, whereas the evaluation of the autocorrelation
function yields exactly the opposite result.
%
%%%\begin{comment}
%
%%%%%%%%%%%%%%%%%%%%%%%%%%%%%%%%%%%%%%%%%%%%%%%%%%%%%%%%%%%%%%%%%%%%%%%%%%%%%%%
%
% Importing a Figure
%
\begin{figure}
\centerline{\psfig{file=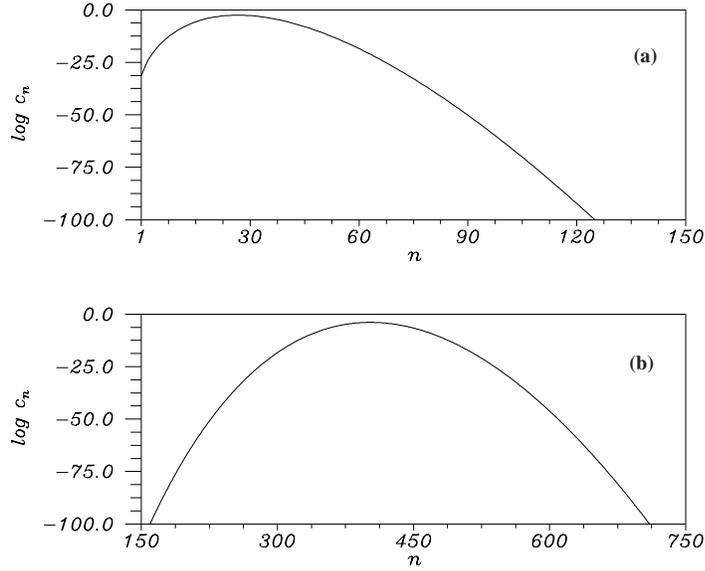,height=7.5cm,angle=0}}
\caption{
The natural logarithm of the coefficients of the autocorrelation function:
In Fig. (a) we set $s=5$ and in Fig. (b) we set $s=20$. 
For a larger $s$ more states yield a nonnegligible 
contribution to the dynamics of the wave packet.
}
\label{k_fig_2}
\end{figure}
%
%%%%%%%%%%%%%%%%%%%%%%%%%%%%%%%%%%%%%%%%%%%%%%%%%%%%%%%%%%%%%%%%%%%%%%%%%%%%%%%
%
%%%\end{comment}
%
This unexpected finding can be understood by further studying the
properties of the coefficients $c_{n}$, and in Fig.'s 2-(a) and 2-(b) we plot
the natural logarithm of the most significant coefficients (times the 
${\rm e}^{ - 2 s^{2} }$ factor) in the expansion of
Eq. (\ref{klauder_8}) for the states of Fig. 1. 
In the state centered around ${\bar n} \approx 25$ only
$\sim 100$ states offer a significant contribution, whereas in the second state 
(${\bar n} \approx 400$) one needs
$\sim 600$ states. Similar results hold for all our other
extensive calculations and the explanation is that
in the limit $s^{2} \gg 1$ the variance
of the distribution of the coefficients $c_{n}$ approaches the Poissonian limit:
\begin{equation}
	{\hspace{-2.1cm}}
	\langle n^{2} \rangle - \langle n \rangle^{2} = 
	\frac{ {\rm e}^{-s^{2}} 
	\left( \frac{ \partial }{ \partial ( s^{2} ) } s^{2} \right)^4 {\rm e }^{s^{2}}
	}{ ( s^{2} + 1 )^{2} + s^{2} }  - 
	\left[
	\frac{ {\rm e}^{-s^{2}} 
	\left( \frac{ \partial }{ \partial ( s^{2} ) } s^{2} \right)^3 {\rm e }^{s^{2}}
	}{ ( s^{2} + 1 )^{2} + s^{2} } 
	\right]^{2} \!
	= s^{2} + 6 + O( \frac{1}{ s^{2} } ) \; .
\label{klauder_11}
\end{equation}
Therefore in the semiclassical limit temporally stable coherent states involve an ever
increasing number of hydrogenic $n$-manifolds, which prevent an accurate
equivalence between the dynamics of wave packets and classical trajectories.

More precisely, the separation of two hydrogenic energy levels with quantum numbers 
${\bar n} \approx s^{2} \gg 1$ and $n = {\bar n} + \Delta$ is:
\begin{equation}
	E_{ n } - E_{\bar{ n}} = 
	\frac{ 2 {\bar n} \Delta + \Delta^{2} }{ 2 {\bar n}^{2} n^{2} }
	= \Delta \frac{ 1 }{ {\bar n}^{3} } - 
	\frac{ 3 \Delta^{2} }{ 2 {\bar n} } \frac{1}{ {\bar n}^{3} }
	+ O \left( \frac{ \Delta^{3} }{ {\bar n}^{2} } \right) 
	\frac{1}{ {\bar n}^{3} } \; .
\label{klauder_12}
\end{equation}
For a narrowly peaked distribution, and in the semiclassical limit the hydrogenic
spectrum is approximately harmonic. That is, the spacings between adjacent 
energy levels are well approximated by an integer multiple of a fundamental
frequency (i.e., $1 / {\bar n}^{3}$), which is the Kepler frequency of the 
eigenstate at the center of the expansion. That is why it is possible to 
construct wave packets which follow classical trajectories with little 
dispersion for at least a few Kepler 
periods {\cite{nauenberg_89a,nauenberg_90a,nauenberg_94a}}.
However, for temporally stable coherent states in the semiclassical limit
the distribution over the eigen-manifolds becomes flatter, and the 
anharmonic corrections of Eq. (\ref{klauder_12}) {\em cannot be neglected}.
For example, 
any term in the autocorrelation function with $n = {\bar n} + \Delta$ and
$\Delta \approx \sqrt{ {\bar n} / 3 }$
is well within the root mean square deviation 
of the distribution ( i.e., $\sqrt{ {\bar n} }$)
and yet after a period $T= 2 \pi {\bar n}^3$ is already
$\approx 180^{\rm{o}}$ out of phase with the ${\bar n}$-term.
Therefore recurrences and even quasi-recurrences become
impossible for time scales comparable to the average Kepler period 
of the wave packet.

\section{Conclusions}

In this paper we have studied the dispersive properties
of Klauder's temporally stable coherent states for the hydrogen atom by calculating
their autocorrelation function. We have found that in the hydrogen 
atom (in contrast with the harmonic oscillator) temporally stable wave packets
do not follow the trajectories of the classical electron remaining well
localized; instead they spread significantly even within a Kepler period.
Such poor quantum-classical equivalence 
does not improve in the semiclassical limit,
i.e., when the expansion of the wave packet is dominated by high-energy eigenstates.
The explanation of this surprising result resides in the 
properties of the distribution over the eigen-manifolds of the 
hydrogen atom, which in the semiclassical limit approaches
a Poisson distribution. 
For large quantum numbers
the variance of the distribution is also very large,
and the anharmonic corrections dominate the dynamics,
so that full or even partial recurrences become impossible.

In the harmonic oscillator temporal stability, coherence and quasi-classical
dynamics are all properties of the same set of states:
this is not true in the hydrogen atom, and 
therefore one should always state clearly the specific properties
which a set of atomic ``coherent states" satisfies.
In fact, in the hydrogen atom
temporally stable coherent states are {\em not} a viable solution
to the problem of finding non-spreading, quasi-classical wave packets, and one has to 
resort either to wave packets which remain localized only for a few Kepler
periods {\cite{nauenberg_89a,nauenberg_94a}} or to modified 
Hamiltonians {\cite{eberly_94a,eberly_96b,delande_95a,delande_95b,turgay_96a,turgay_97a}} 
in which external fields bring about the 
desired localization of the wave packet.

\section*{References}

\end{document}